\documentclass[aps,pra,reprint,groupedaddress,showpacs,floatfix]{revtex4-1}

\usepackage[dvips]{graphicx}  
\usepackage{amsmath,amssymb,bm}
\usepackage{dcolumn}

\begin{document}


\title{Dark-state suppression and optimization of laser cooling and fluorescence in a trapped alkaline-earth-metal single ion}


\author{T.\ Lindvall}
\email[]{thomas.lindvall@mikes.fi}
\author{M.\ Merimaa}
\affiliation{Centre for Metrology and Accreditation (MIKES), P.O.\ Box 9, FI-02151 Espoo, Finland}
\author{I.\ Tittonen}
\affiliation{Department of Micro- and Nanosciences, Aalto University, P.O.\ Box 13500, FI-00076 Aalto, Finland}
\author{A.~A.\ Madej}
\affiliation{Frequency and Time Group, Measurement Science and Standards Portfolio, National Research Council of Canada, Ottawa, Ontario K1A 0R6, Canada}


\date{\today}

\begin{abstract}
We study the formation and destabilization of dark states in a single trapped $^{88}\mathrm{Sr}^+$ ion caused by the cooling and repumping laser fields required for Doppler cooling and fluorescence detection of the ion. By numerically solving the time-dependent density matrix equations for the eight-level system consisting of the sublevels of the $5s\,^2S_{1/2}$, $5p\,^2P_{1/2}$, and $4d\,^2D_{3/2}$ states, we analyze the different
types of dark states and how to prevent them in order to maximize the scattering rate, which is crucial for both the cooling and the detection of the ion.
The influence of the laser linewidths and ion motion on the scattering rate and the dark resonances is studied.
The calculations are then compared with experimental results obtained with an endcap ion trap system located at the National Research Council of Canada and found to be in good agreement.
The results are applicable also to other alkaline earth ions and isotopes without hyperfine structure.
\end{abstract}


\maketitle


\section{Introduction}

A single ion in a radiofrequency (RF) ion trap can be laser cooled to a low temperature and confined to a region in space with dimensions less than an optical wavelength. It thus constitutes a pure and well isolated quantum system that lends itself to applications where control and isolation from the environment are called for: Narrow optical transitions in trapped ions can be used to realize optical frequency standards \cite{Madej2001a,Margolis2009a}, and ion traps are also promising systems for the implementation of a scalable quantum computer \cite{Leibfried2003a,Wineland2011a}.

\begin{figure}[t]
\includegraphics[width=0.67\columnwidth]{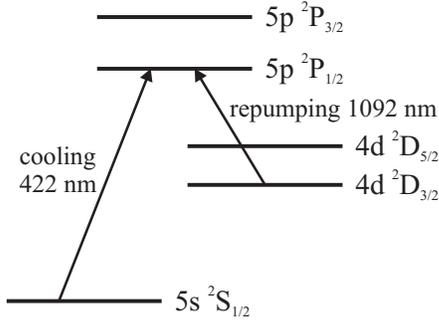}%
\caption{\label{fig:SrII}Lowest energy levels of $^{88}\mathrm{Sr}^+$. Cooling and repumping transitions are shown as arrows together with their respective wavelengths.}
\end{figure}

We study even isotopes of the alkaline-earth-metal ions that have no nuclear spin and thus no hyperfine structure, most notably $^{24}$Mg$^+$, $^{40}$Ca$^+$, $^{88}$Sr$^+$, and $^{138}$Ba$^+$, but also including less abundant isotopes of these elements. Numerical calculations and a comparison with experimental results are performed for $^{88}\mathrm{Sr}^+$, the lowest-lying energy levels of which are shown in Fig.~\ref{fig:SrII}. The $^2S_{1/2} - \,^2P_{1/2}$ transition is used to Doppler-cool the ion. The narrow $^2S_{1/2} - \,^2D_{5/2}$ electric quadrupole transition is used as the clock transition in optical clocks, for sub-Doppler cooling using resolved sideband cooling \cite{Wineland1975a}, and as a qubit candidate for quantum computing \cite{Nagerl2000a}. The occurrence of a transition to the $^2D_{5/2}$ state is detected using Dehmelt's electron shelving technique \cite{Dehmelt1982a,Dehmelt1975a}, i.e., as a dark period in the fluorescence caused by the cooling laser. The $^2P_{1/2}$ excited state has a finite probability of decaying to the metastable $^2D_{3/2}$ state. This requires a repumping laser tuned to the $^2D_{3/2} - \,^2P_{1/2}$ transition that returns the ion to the cooling cycle.

A high scattering rate is crucial for both Doppler cooling and fluorescence detection of the trapped ion. It is compromised by the formation of dark states, which can be angular momentum eigenstates or superpositions of Zeeman sublevels of the $^2S_{1/2}$ and/or $^2D_{3/2}$ states (coherent population trapping, CPT \cite{Alzetta1976a,Arimondo1996a}). CPT resonances between the $^2S_{1/2}$ and the $^2D_{3/2}$ states can be experimentally observed by sweeping the detuning of one of the laser beams  \cite{Janik1985a,Klein1990a,Siemers1992a,Kurth1995a,Barwood1998a}. It is also known that optical pumping into dark superposition states between the $^2D_{3/2}$ sublevels can occur when the magnetic field is weak and that this can be prevented for example by modulating the polarization of the repumper \cite{Barwood1998a,Berkeland1998b}.

Berkeland and Boshier \cite{Berkeland2002a} have theoretically studied how to destabilize dark states in Zeeman degenerate systems using an external magnetic field or by modulating the laser polarization. Both methods rely on making the dark states evolve in time more rapidly than the optical pumping processes.
One of the systems they consider is the $^2S_{1/2} - \,^2P_{1/2} - \,^2D_{3/2}$ system. They recommend Rabi frequencies that give a reasonable scattering rate without excessive power broadening and point out that the repumper should have a positive detuning to tune the system away from CPT resonance (the cooling laser is red-detuned for Doppler cooling). The magnetic field or rate of polarization modulation required to destabilize the dark states is then analyzed.

This paper expands on the results in Ref.~\cite{Berkeland2002a}. We allow the two laser fields to have different polarizations and take into account the effect of the laser linewidths and correlation properties on the CPT resonances. Furthermore, the effect of the motion of the ion, secular and micromotion, is analyzed. Finally, the theoretical results are compared to experimental spectra from the $^{88}\mathrm{Sr}^+$ endcap trap at the National Research Council of Canada (NRC) \cite{Dube2010a}.

\section{Theory}

\begin{figure}
\includegraphics[width=1\columnwidth]{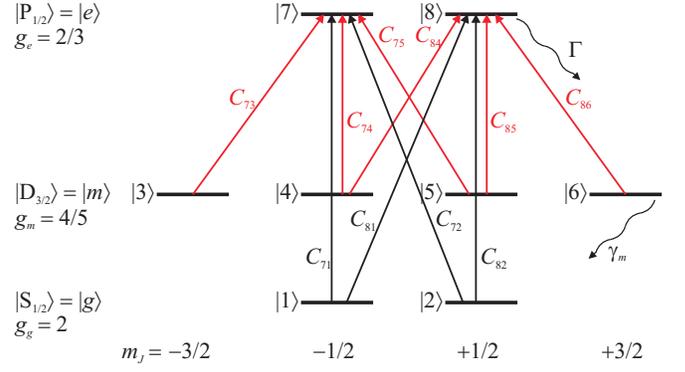}%
\caption{\label{fig:8-level}(Color online) The eight-level system. Fine-structure state designations are given at the left together with the respective Land\'e $g$ factors. Magnetic quantum numbers are shown at the bottom. Sublevels are labeled consecutively and the allowed transitions between them are shown as arrows together with the corresponding relative transition amplitudes $C_{ij}$.}
\end{figure}

The atomic Hamiltonian of the eight-level system shown in Fig.~\ref{fig:8-level} is

\begin{equation}
H_\mathrm{a} = \hbar\omega_g \sum_{i=1}^2 |i\rangle\langle i| + \hbar\omega_m \sum_{i=3}^6 |i\rangle\langle i| + \hbar\omega_e \sum_{i=7}^8 |i\rangle\langle i|,
\end{equation}
where $\omega_g$, $\omega_m$, and $\omega_e$ are the frequencies of the ground state $|g\rangle$, metastable state $|m\rangle$, and excited state $|e\rangle$, respectively.
The interaction with an external magnetic field $\mathbf{B}$, $V_\mathrm{m} = - \bm{\mu}_\mathrm{m} \cdot\mathbf{B}$, where $\bm{\mu}_\mathrm{m}$ is the magnetic dipole moment, is included in the atomic Hamiltonian in the form of (linearly) Zeeman shifted levels,
\begin{equation}
H_\mathrm{a,m} = \sum_{i=1}^8 \hbar\omega_i |i\rangle\langle i|.
\end{equation}
Here $\omega_i = \omega_l + g_l m_{J,i} \mu_\mathrm{B} B/\hbar$, where $l$ is $g$, $m$, or $e$ and $g_l$ is the corresponding Land\'e factor, $m_{J,i}$ is the magnetic quantum number of the level $|i\rangle$, and $\mu_\mathrm{B}$ is the Bohr magneton. We have chosen the quantization axis (QA), i.e., the $z$ axis, along the magnetic field. The ion is located at the origin. It is first assumed to be stationary, later its motion is accounted for.

The system is interacting with two laser fields that can have arbitrary time-dependent polarizations. The cooling field is slightly red-detuned from the $|g\rangle \rightarrow |e\rangle$ transition (for Doppler cooling) and induces the transitions shown as black arrows in Fig.~\ref{fig:8-level}, while the repumping field is near resonant with the $|m\rangle \rightarrow |e\rangle$ transition and induces the transitions shown as gray (red) arrows in Fig.~\ref{fig:8-level}. The electric field of the laser field $i=\mathrm{c}, \mathrm{r}$ is written as
\begin{equation} \label{eq:E_laser}
\mathbf{E}_i(t) = \frac{\mathcal{E}_i}{2} \left[ f_0^i(t) \mathbf{u}_0 + f_{+1}^i(t) \mathbf{u}_{+1} + f_{-1}^i(t) \mathbf{u}_{-1}\right] e^{-i\omega_i t} + \text{c.c.},
\end{equation}
where $\mathcal{E}_i$ is the electric field amplitude, $\omega_i$ is the frequency, c.c.\ stands for complex conjugate and $f_q^i(t)$ is the time-dependent complex amplitude of the polarization component $q$, obtained as the projection of the unit polarization vector of the field onto the spherical unit vector $\mathbf{u}_q$. These are
\begin{equation}
\mathbf{u}_{\pm1} = \mp \frac{1}{\sqrt{2}} (\mathbf{u}_x \pm i\mathbf{u}_y), \quad \mathbf{u}_{0} = \mathbf{u}_z,
\end{equation}
and describe $\sigma^\pm$ and $\pi$ polarized light, respectively.

In the electric dipole and rotating wave approximations, the interaction Hamiltonians for the two fields become
\begin{eqnarray}
V_\mathrm{c}(t) = -\frac{\hbar \Omega_\mathrm{c}}{2} \big[&f^\mathrm{c}_0(t)& (C_{71}|7\rangle\langle 1| + C_{82}|8\rangle\langle 2|) \nonumber \\*
&+& f^\mathrm{c}_{+1}(t) C_{81}|8\rangle\langle 1| \nonumber \\*
&+& f^\mathrm{c}_{-1}(t) C_{72}|7\rangle\langle 2|\big] e^{-i\omega_\mathrm{c}t} + \text{H.c.}
\end{eqnarray}
and
\begin{eqnarray}
V_\mathrm{r}(t) = -\frac{\hbar \Omega_\mathrm{r}}{2} \big[&f^\mathrm{r}_0(t)& (C_{74}|7\rangle\langle 4| + C_{85}|8\rangle\langle 5|) \nonumber \\*
&+& f^\mathrm{r}_{+1}(t) (C_{73}|7\rangle\langle 3| + C_{84}|8\rangle\langle 4|) \nonumber \\*
&+& f^\mathrm{r}_{-1}(t) (C_{75}|7\rangle\langle 5|+C_{86}|8\rangle\langle 6|)\big] \nonumber \\*
&\times& e^{-i\omega_\mathrm{r}t} + \text{H.c.},
\end{eqnarray}
where H.c.\ stands for Hermitian conjugate. The coefficients $C_{ji}$ are the relative transition amplitudes of the electric dipole transitions and can be obtained using the Wigner-Eckart theorem, which states how a matrix element depends on the $m_J$ quantum numbers~\cite{Sobelman:ASRT}
\begin{eqnarray} \label{eq:W-E_theorem}
\langle J' m_J' | T^\kappa_q |J m_J\rangle &=& (-1)^{J'-m_J'} \langle J'||T^\kappa ||J\rangle
\begin{pmatrix} J' & \kappa & J \\ -m_J' & q & m_J \end{pmatrix} \nonumber \\
&=& C_{J'm_J',Jm_J} T^\kappa_{J',J}.
\end{eqnarray}
Here $T^\kappa_q$ is a set of tensor operators of rank $\kappa$ and $T^\kappa_{J',J} = \langle J'||T^\kappa ||J\rangle$ is the reduced matrix element that does not depend on $m_J$, $m_J'$, or $q$. Using Eq.~(\ref{eq:W-E_theorem}) for the electric dipole operator ($\kappa=1$; $q = 0$ for $\pi$ and $\pm1$ for $\sigma^\pm$ transitions), one obtains: $C_{82}=-C_{71}=6^{-1/2}$, $C_{72}=-C_{81}=3^{-1/2}$, $C_{73}=C_{86}=2^{-1}$, $C_{74}=C_{85}=-6^{-1/2}$, and $C_{75}=C_{84}=2^{-1}3^{-1/2}$.

The ``two-level'' Rabi frequencies corresponding to the two laser fields are $\Omega_\mathrm{c} = \mu_{P_{1/2},S_{1/2}} \mathcal{E}_\mathrm{c} /\hbar$ and $\Omega_\mathrm{r} = \mu_{P_{1/2},D_{3/2}} \mathcal{E}_\mathrm{r} /\hbar$, where $\mu_{L'_{J'},L_J} = \langle L'_{J'}||-e\mathbf{r} ||L_J\rangle$ is the reduced dipole moment. The Rabi frequencies can be assumed to be real. The true Rabi frequency for a certain transition $|i\rangle \rightarrow |j\rangle$ is equal to $C_{ji} \Omega_k$ ($k=c,r$ depending on which laser field drives the transition).
Note that Ref.~\cite{Berkeland2002a} uses rms Rabi frequencies defined as $\Omega_\text{rms}^2 = \sum_{m_J,m_J'} |\Omega_{m_J m_J'}|^2$. From the sum rule \cite{Sobelman:ASRT}
\begin{equation} \label{eq:matrix_sum}
\sum_{m_J, m_J'} |\langle J' m_J' | T^\kappa_q |J m_J\rangle|^2 = \frac{1}{2\kappa +1} |T^\kappa_{J',J}|^2,
\end{equation}
we obtain the relation between the rms and the two-level Rabi frequencies: $\Omega_\text{rms} = 3^{-1/2} \Omega$.

The Liouville-von Neumann equation for the density matrix of the eight-level system is
\begin{equation} \label{eq:L-vN}
\frac{\mathrm{d}\rho}{\mathrm{d}t} = \frac{1}{i\hbar}[H_\mathrm{a,m}+V_\mathrm{c}(t)+V_\mathrm{r}(t),\rho] +\frac{\mathrm{d}\rho^\text{relax}}{\mathrm{d}t}.
\end{equation}
Terms oscillating at optical frequencies are eliminated by the substitutions
\begin{subequations}
\begin{eqnarray}
\tilde{\rho}_{kl} &=& \rho_{kl} e^{-i\omega_\mathrm{c}t}, \quad 1\leq k\leq 2, \;  7\leq l\leq 8, \\
\tilde{\rho}_{kl} &=& \rho_{kl} e^{-i\omega_\mathrm{r}t}, \quad 3\leq k\leq 6, \;  7\leq l\leq 8, \\
\tilde{\rho}_{kl} &=& \rho_{kl} e^{-i(\omega_\mathrm{c}-\omega_\mathrm{r})t}, \quad 1\leq k\leq 2, \;  3\leq l\leq 6.
\end{eqnarray}
\end{subequations}

\subsection{Relaxation}

The relaxation terms $\mathrm{d}\rho^\text{relax}/\mathrm{d}t$ in Eq.~(\ref{eq:L-vN}) are added phenomenologically. The excited state populations decay at the rate $\Gamma$. The excited state decays to the ground and metastable states with the probabilities $A_g$ and $A_m=1-A_g$ and the decay probability for a certain magnetic sublevel transition, e.g., $|8\rangle$ to $|6\rangle$ can then be calculated as
\begin{equation}
A_{86} = \frac{|C_{86}|^2 A_m}{|C_{86}|^2+|C_{85}|^2+|C_{84}|^2}.
\end{equation}

The excited state coherence $\rho_{78}$ decays at the rate $\Gamma$ and can be partially transferred to the ground state coherences in the spontaneous emission process~\cite{Cohen-Tannoudji1977a}
\begin{eqnarray}
\frac{\mathrm{d}\rho_{m_g,m_g'}^\text{relax}}{\mathrm{d}t} = \Gamma_{e\rightarrow g} (2J_e+1)
\sum_{q=-1}^1 C_{J_e m_e, J_g m_g} C_{J_e m_e', J_g m_g'} \nonumber \\
\times \rho_{m_e=m_g+q,m_e'=m_g'+q}, \nonumber \\
 { }
\end{eqnarray}
where the transition amplitudes $C_{J_e m_e, J_g m_g}$ are as defined above. This adds the following relaxation terms to the evolution equations
\begin{subequations}
\begin{eqnarray}
\frac{\mathrm{d}\rho_{i,i+1}^\text{relax}}{\mathrm{d}t} &=& 2 C_{7i} C_{8,i+1} A_m \Gamma \rho_{78}, \quad i = 3,4,5, \\
\frac{\mathrm{d}\rho_{12}^\text{relax}}{\mathrm{d}t} &=& 2 C_{71} C_{82} A_g \Gamma \rho_{78}.
\end{eqnarray}
\end{subequations}

The optical coherences decay at the rate $\Gamma/2$ (we assume the laser linewidths to be $\ll\Gamma$) and the metastable state populations and coherences decay at the rate $\gamma_m \ll \Gamma$. The transition probabilities for the quadrupole transitions can be obtained from Eq.~\ref{eq:W-E_theorem} with $\kappa = 2$ and the branching ratios for decay from the metastable state are then $A_{mg} = |C_{mg}|^2/\sum_g  |C_{mg}|^2$: $A_{62}=A_{31}=1/5$, $A_{61}=A_{32}=4/5$, $A_{52}=A_{41}=2/5$, and $A_{51}=A_{42}=3/5$.

We have assumed that there are no external processes, such as background gas collisions, that cause transitions between the ground or metastable $m_J$ levels or dephasing of the ground and metastable coherences. The dephasing rate of the coherences between the ground and the metastable levels depends on the laser linewidths \cite{Arimondo1996a}
\begin{equation} \label{eq:g-m_dephasing}
\gamma_{g,m} = \frac{1}{2}\left( \gamma_m + \Delta\omega_\mathrm{c}+\Delta\omega_\mathrm{r} - 2\Delta\omega_\mathrm{c,r} \right),
\end{equation}
where $\Delta\omega_\mathrm{c}$ and $\Delta\omega_\mathrm{r}$ are the FWHM linewidths of the cooling and repumper lasers and $\Delta\omega_\mathrm{c,r}$ is the cross-correlated linewidth. For perfectly cross-correlated lasers the linewidths cancel, whereas for completely uncorrelated lasers $\Delta\omega_\mathrm{c,r} = 0$ and the linewidths add up. In the low-intensity limit, the linewidth of the CPT resonances is $2\gamma_{g,m}$.

\subsection{Solution of density matrix equations}

The time-dependent density matrix is solved by direct numerical integration of Eq.~(\ref{eq:L-vN}) with the initial condition $\rho_{11}=\rho_{22}=1/2$, i.e., a thermally populated ground state. This describes the situation well when the cooling and repumping beams are turned on after a probe-laser interrogation on the $^2S_{1/2} - \,^2D_{5/2}$ quadrupole transition. When the ion is first trapped,  the situation is more complicated, as it is hot and there might be another far-detuned cooling field present \cite{Brownnutt2007a}. However, also in this case the treatment should be valid for the ``final'' state, i.e., when the ion has cooled down and reached quasi-steady state (when the only time dependence is due to the possible polarization modulation of the repumper). As the density matrix describes ensemble averages, the time-dependent solution should be interpreted as the average over an infinite number of experimental cycles for a single trapped ion.

If the polarization amplitudes $f_q^i$ are time independent, the steady-state solution can be solved by setting the time derivative to 0 in Eq.~(\ref{eq:L-vN}). If they are time-dependent, we must integrate Eq.~(\ref{eq:L-vN}) until quasi-steady state is reached and average the density matrix over one polarization modulation period, which is considerably more computer time consuming.

The (quasi-)steady-state value of the scattering rate is the most relevant parameter for comparison with experiments. To collect the fluorescence of the single ion, it is customary to use a bandpass filter at the cooling wavelength to minimize problems with background light and scattering. We therefore define the scattering rate as $\Gamma_\mathrm{sc} = A_g \Gamma (\rho_{77}+\rho_{88})$.

\section{Experimental parameters for $^{88}$S\MakeLowercase{r}$^+$ \label{sec:exppar}}

Table \ref{tab:exp_par} summarizes the numerical parameters used for the $^{88}$Sr$^+$ ion in this paper. We use the dipole matrix elements, spontaneous decay rate, and branching ratio calculated by Jiang \emph{et al.}~\cite{JiangD2009a}. Their values are in good agreement with the experimental $^2 P_{1/2}$ lifetime in \cite{Pinnington1995a} and with the calculated lifetime and oscillator strengths in \cite{Brage1998a}. Gallagher's~\cite{Gallagher1967a} experimental oscillator strengths and branching ratio are often cited. However, his oscillator strength for the $^2S_{1/2} - \,^2P_{1/2}$ transition agrees with the results in \cite{JiangD2009a}, but he acknowledges that his value for the $^2D_{3/2} - \,^2P_{1/2}$ transition probably is too large, which is why we chose to use the values of Jiang \emph{et al.}~\cite{JiangD2009a}.

\begin{table}
\caption{Numerical parameters for $^{88}$Sr$^+$.\label{tab:exp_par}}
\begin{ruledtabular}
\begin{tabular}{l d c}
Parameter & \multicolumn{1}{c}{Value} & Ref.\ No. \\ \hline
$\mu_{P_{1/2},S_{1/2}}$ & 3.078 \,ea_0 & \cite{JiangD2009a} \\
$\mu_{P_{1/2},D_{3/2}}$ & 3.112 \,ea_0 & \cite{JiangD2009a} \\
$\Gamma$ & 135.58 \times 10^{6}\;\mathrm{s}^{-1} & \cite{JiangD2009a} \\
$A_g$ & 0.9444  & \cite{JiangD2009a} \\
$\gamma_m$ & 2.30\;\mathrm{s}^{-1}  & \cite{Mannervik1999a} \\
\end{tabular}
\end{ruledtabular}
\end{table}

Using $I_i = \varepsilon_0 c \mathcal{E}_i^2 /2$, we can relate the two-level Rabi frequencies to the laser intensities
\begin{equation} \label{eq:Rabi_int}
    \Omega_i = \frac{\mu_i}{\hbar}\sqrt{\frac{2I_i}{\varepsilon_0 c}}, \quad i=\mathrm{c},\mathrm{r},
\end{equation}
where $\mu_i$ is shorthand notation for the corresponding reduced dipole moment. This gives the numerical relations $\Omega_\mathrm{c}/\Gamma = (I_\mathrm{c}/39.8\;\text{mW}\text{cm}^{-2})^{1/2}$ and $\Omega_\mathrm{r}/\Gamma = (I_\mathrm{r}/39.0\;\text{mW}\text{cm}^{-2})^{1/2}$.
The laser beams can be carefully adjusted, so we assume that the ion experiences the peak intensity of the Gaussian beam, $I_i=2P_i/\pi w_i^2$, where $P_i$ is the power and $w_i$ is the waist.

Berkeland and Boshier \cite{Berkeland2002a} recommend using the two-level Rabi frequencies $\Omega_\mathrm{c} \approx \Omega_\mathrm{r} \approx 3^{-1/2}\Gamma$ in order to obtain a reasonable scattering rate with little power broadening. However, considerably higher cooling laser Rabi frequencies have been used in experiments at both the NRC ($\Omega_\mathrm{c} \approx 20\Gamma$) \cite{Madej2004a} and the National Physical Laboratory ($\Omega_\mathrm{c} \approx 7\Gamma$) \cite{Brownnutt2007}.

\section{Types of dark states \label{sec:dark_states}}

\subsection{Ground-level dark states \label{sec:g-states}}

Berkeland and Boshier~\cite{Berkeland2002a} state that for a transition between states with half-integer angular momentum and $J_g = J_e$, such as the cooling transition considered here, there is one dark state for circular polarization only. Let us elaborate on this.

\begin{figure}[h]
\includegraphics[width=.85\columnwidth]{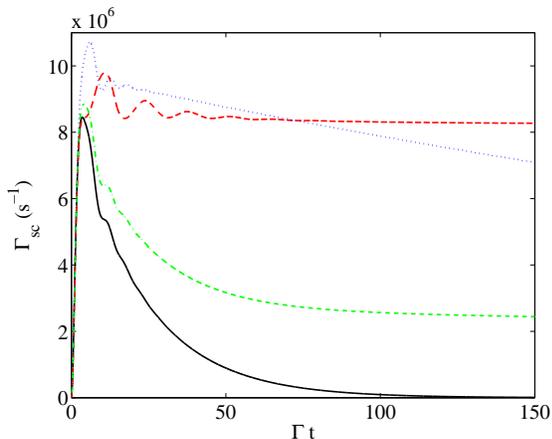}%
\caption{\label{fig:g-states}(Color online) Time-dependent scattering rates for different cooling polarizations. A $\sigma^+$ polarized cooling beam rapidly pumps the population into the dark state $|2\rangle$ [$B=0$; solid (black) curve]. In a transversal magnetic field, the dark state is a superposition state, which is destabilized by the field [$\mathbf{B} = 350\;\mu\mathrm{T} \mathbf{u}_x$; dashed (red) curve]. A combination of $\pi$ and $\sigma^+$ polarization creates a partially dark state [$B=0$; dash-dotted (green) curve]. For a linearly polarized cooling beam, there are no ground-level dark states [$B=0$; dotted (blue) curve]. For all curves, $\Omega_\mathrm{c} = \Omega_\mathrm{r} = \Gamma$, $\delta_\mathrm{c} = -\Gamma/2$, $\delta_\mathrm{r} = \Gamma/2$, $\gamma_{g,m}=\Gamma/10$, and repumper polarized along $\mathbf{u}_y$.}
\end{figure}

Figure \ref{fig:g-states} shows the scattering rate as a function of time for different cooling laser polarizations. For a circularly polarized cooling beam in zero magnetic field, we can choose the QA such that the polarization is $\sigma^+$. This polarization only drives the $|1\rangle \rightarrow |8\rangle$ transition and the scattering rate decays rapidly as the population is pumped into the dark state $|2\rangle$ at the optical pumping rate $\Gamma_\mathrm{op} = 0.045 \Gamma$ [solid (black) curve in Fig.~\ref{fig:g-states}]. We have earlier derived an expression for the optical pumping rate in a three-level system with one dark lower level~\cite{Lindvall2009a}
\begin{equation} \label{eq:Gamma_op}
\Gamma_\mathrm{op} = \frac{A_n \Gamma \left(\frac{C\Omega}{2}\right)^2}{\delta^2 + \left(\frac{\Gamma}{2}\right)^2 + \left[ 2+\frac{\delta^2-3 \left(\Gamma/2 \right)^2}{\delta^2 + \left( \Gamma/2 \right)^2} A_n \right] \left(\frac{C\Omega}{2}\right)^2}.
\end{equation}
Here $A_n$ is the spontaneous decay probability to the noncoupled state. If we neglect the metastable state, Eq.~(\ref{eq:Gamma_op}) can be applied to the effective three-level system consisting of $|1\rangle$, $|2\rangle$, and $|8\rangle$ ($A_n=A_{82}$, $C=C_{81}$), giving $\Gamma_\mathrm{op} = 0.041 \Gamma$ in reasonable agreement with the numerical result above. If we apply a magnetic field in the direction of the cooling beam, $|2\rangle$ is still a dark eigenstate and the Zeeman shift only changes the detuning.

If we choose the QA in another direction, e.g., in the $x$ direction when the beam is in the $z$ direction, the polarization in the QA frame is $-2^{-1/2}[\mathbf{u}_0-i2^{-1/2}(\mathbf{u}_{+1} - \mathbf{u}_{-1})]$, i.e., a combination of $\pi$ and $\sigma$ ($\sigma$ refers to light that has a linear polarization orthogonal to the QA and that thus induces $\sigma^+$ and $\sigma^-$ transitions). The dark state is now the superposition state
\begin{equation}
|\text{DS}_{1,2}\rangle = \frac{1}{\sqrt{2}} (|1\rangle - i|2\rangle),
\end{equation}
for which $\langle 7 |V_\mathrm{c}|\text{DS}_{1,2}\rangle = \langle 8 |V_\mathrm{c}|\text{DS}_{1,2}\rangle = 0$.
If we apply a magnetic field in the direction of the QA ($\mathbf{u}_x$), $|\text{DS}_{1,2}\rangle$ is no longer an energy eigenstate, but the dark state evolves in time
\begin{equation} \label{eq:DStimedep}
|\text{DS}_{1,2}\rangle (t) = \frac{e^{-i\omega_gt}}{\sqrt{2}}
(e^{ig_g\mu_\mathrm{B}Bt/2\hbar}|1\rangle - i e^{-ig_g\mu_\mathrm{B}Bt/2\hbar}|2\rangle).
\end{equation}
When the field is strong enough that the evolution rate of the dark state $g_g \mu_\mathrm{B} B/\hbar$ is higher than the optical pumping rate $\Gamma_\mathrm{op}$ (corresponding to $B\approx 35\;\mu$T for the parameters in Fig.~\ref{fig:g-states}), the system cannot follow and the dark state is destabilized. Another way to explain this is that the optical pumping rate $\Gamma_\mathrm{op}$ determines the width of the ground states, and the Zeeman shift of the ground levels tune the $\Lambda$ system out of Raman (CPT) resonance. When $g_g \mu_\mathrm{B} B/\hbar \approx 10 \Gamma_\mathrm{op}$ ($B\approx 350\;\mu$T), the dark state is completely destabilized; see the dashed (red) curve in Fig.~\ref{fig:g-states}. If the field is increased further so that the Zeeman shifts approach $\Gamma$, the cooling laser is tuned away from resonance and the scattering rate decreases.

If the polarization is a combination of $\pi$ and $\sigma^+$ (this can be achieved by an elliptically polarized beam traveling at an angle to the QA), the $\Lambda$ system $|1\rangle \leftrightarrow |8\rangle \leftrightarrow |2\rangle$ is formed, but it is not closed, as the $|1\rangle \leftrightarrow |7\rangle$ transition is also driven. Hence the contrast of the corresponding CPT resonance is less than unity, i.e., the scattering rate decays rapidly, but not to 0 [dash-dotted (green) curve in Fig.~\ref{fig:g-states}].

If the cooling laser is linearly polarized ($\pi$, $\sigma$, or combination thereof in the QA frame), there is no dark state among the ground levels. The slow decay of the dotted (blue) curve in Fig.~\ref{fig:g-states} is due to dark states among the metastable sublevels, which are considered in Sec.~\ref{sec:m-states}. In the following, we therefore only consider a linearly polarized cooling beam.

\subsection{Ground-metastable-level dark states}

The ground-metastable-level dark states are CPT superposition states of $|g\rangle$ and $|m\rangle$ sublevels. As the two legs of the $\Lambda$ systems are driven by separate lasers, the CPT resonances can be detected by tuning one of the lasers over the Raman resonance \cite{Janik1985a,Klein1990a,Siemers1992a,Kurth1995a,Barwood1998a}.

The width and contrast of the ground-metastable CPT resonances depend strongly on the ground-metastable  coherence dephasing rate $\gamma_{g,m}$, Eq.~(\ref{eq:g-m_dephasing}). Siemers \emph{et al}.\ \cite{Siemers1992a} used  narrow-band lasers (``a few kHz'') and deduced a coherence dephasing rate of $2\pi\times 12$\;kHz, in agreement with the assumption of noncorrelated lasers (the actual CPT resonances were wider due to power broadening).

\begin{figure}[h]
\includegraphics[width=.85\columnwidth]{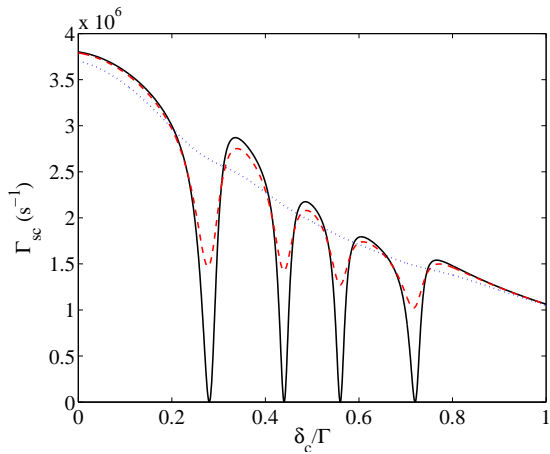}%
\caption{\label{fig:gammaL}(Color online) CPT resonances for different values of $\gamma_{g,m}/\Gamma$: 0 [solid (black) curve], 0.01 [dashed (red) curve], and 0.1 [dotted (blue) curve]. $\Omega_\mathrm{c} = \Omega_\mathrm{r} = \Gamma/2$, $\delta_\mathrm{r} = \Gamma/2$, $\mu_\mathrm{B} B/\hbar = \Gamma/10$ ($B\approx 150\;\mu$T).}
\end{figure}

\begin{figure}[h]
\includegraphics[width=.85\columnwidth]{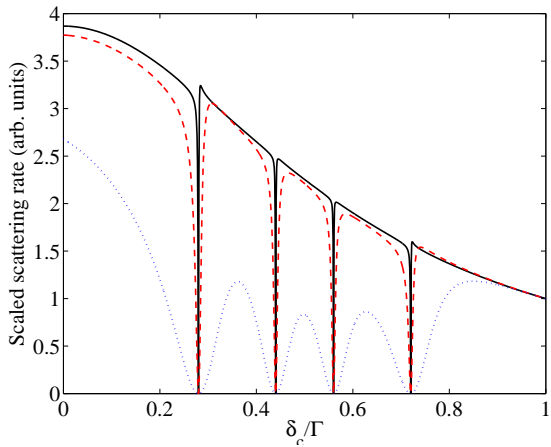}%
\caption{\label{fig:powerbroadening}(Color online) Power broadening of the CPT resonances. Normalized (at $\delta_\mathrm{c}=\Gamma$) scattering rate spectra for different Rabi frequencies $\Omega_\mathrm{c} = \Omega_\mathrm{r}$: $0.1\Gamma$ [solid (black) curve], $0.3\Gamma$ [dashed (red) curve], and $\Gamma$ [dotted (blue) curvee]. $\gamma_{g,m}=0$, $\delta_\mathrm{r} = \Gamma/2$, $\mu_\mathrm{B} B/\hbar = \Gamma/10$ ($B\approx 150\;\mu$T).}
\end{figure}

The dephasing due to the laser linewidths was not examined in the calculations in Ref.~\cite{Berkeland2002a}. Figure~\ref{fig:gammaL} shows ground-metastable CPT resonances for parameters corresponding to curve A in Fig.~7 of Ref.~\cite{Berkeland2002a}, but for different dephasing rates $\gamma_{g,m}$. As this rate approaches $\Gamma$, the CPT resonances disappear completely. The magnetic field of $150\;\mu$T is large enough to separate the four resonances as well as to destabilize metastable-level dark states (see Sec.~\ref{sec:m-states}). The CPT resonances have finite linewidths even for  $\gamma_{g,m}=0$ due to power broadening, which is illustrated in Fig.~\ref{fig:powerbroadening}.
Note that the blue side of the $^2S_{1/2} - ^2\,P_{1/2}$ transition cannot be scanned in a real ion trap experiment as the ion is lost due to heating.

As the cooling laser is red-detuned (typically $\delta_c\approx -\Gamma/2$), ground-metastable CPT resonances can be avoided by tuning the repumper slightly to the blue. If the repumper is modulated so that sidebands are created, care must be taken that no sideband is at Raman resonance either, as discussed in Sec.~\ref{sec:polarizationmodulation}.

\subsection{Metastable-level dark states \label{sec:m-states}}

\begin{figure}[h]
\includegraphics[width=.85\columnwidth]{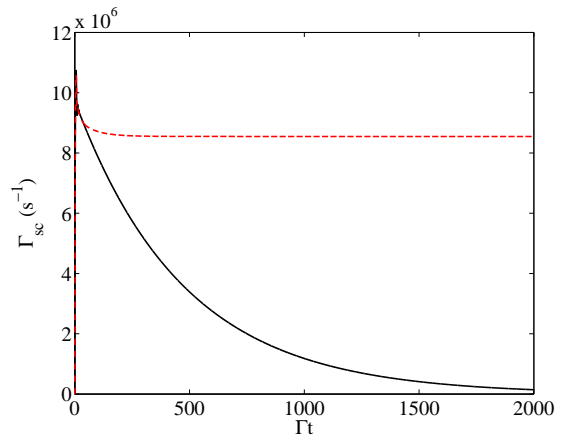}%
\caption{\label{fig:m-states}(Color online)  For $B=0$, the scattering rate is independent of the repumper polarization [$\pi$, $\sigma$, or $\sigma^+$; solid (black) curve]. For a $\sigma$ polarized repumper, a field of 100\;$\mu$T ($2 g_m \mu_\mathrm{B} B/\hbar \approx 0.09 \Gamma$) destabilizes the dark states completely [dashed (red) curve]. $\Omega_\mathrm{c} = \Omega_\mathrm{r} = \Gamma$, $\delta_\mathrm{c} = -\Gamma/2$, $\delta_\mathrm{r} = \Gamma/2$, $\gamma_{g,m}=\Gamma/10$, and cooling laser $\sigma$ polarized.}
\end{figure}

The $|J_m=3/2\rangle \rightarrow |J_e=1/2\rangle$ repumper transition has dark states for all laser polarizations. In zero field, the scattering rate is practically independent of the repumper polarization [see the solid (black) curve in Fig.~\ref{fig:m-states}]. The optical pumping rate is  $\Gamma_\mathrm{op} \approx 0.002 \Gamma$ and again Eq.~(\ref{eq:Gamma_op}) gives a good estimate if we assume both fields to be linearly polarized so that we can use $C=C_{82}$ and $A_n=A_{86}$. This optical pumping rate is lower than the optical pumping rate into ground-level dark states (Sec.~\ref{sec:g-states}) by roughly a factor of $A_m/A_g$ as one would expect. The actual dark states depend on the repumper polarization. For $\pi$ polarization, $|3\rangle$ and $|6\rangle$ are dark states, for $\sigma^+$, $|5\rangle$ and $|6\rangle$ are dark, and for $\sigma$ polarization, the dark states are superpositions of $|3\rangle$ and $|5\rangle$ and of $|4\rangle$ and $|6\rangle$.

When the repumper is $\sigma$ polarized, the dark states can again be destabilized using a magnetic field. The relation between the optical pumping rate and the required field is not as simple as in Sec.~\ref{sec:g-states} because the decay probability $A_m$ is low. However, a field of 100\;$\mu$T (dark state evolution rate $2 g_m \mu_\mathrm{B} B/\hbar \approx 0.09 \Gamma$) is enough to fully destabilize the dark states for the parameters in Fig.~\ref{fig:m-states} [dashed (red) curve].

\section{Polarization angles}

Let us summarize the requirements on the laser polarizations. In Sec.~\ref{sec:g-states} we showed that the cooling laser beam should be linearly polarized. If we are far from Raman resonance, its function, from the point of view of dark states, is merely to pump population into the metastable levels, where the repumper can then build up coherences (dark states) unless these are destabilized. Hence the effect of the angle between the magnetic field and the linear cooling laser polarization is negligible, typically only a few percent for the parameters considered here. On the other hand, near Raman resonance the number, position, and amplitude of the ground-metastable CPT resonances depend strongly on the polarization of both laser fields, but this is a regime we try to avoid.

\begin{figure}[h]
\includegraphics[width=.85\columnwidth]{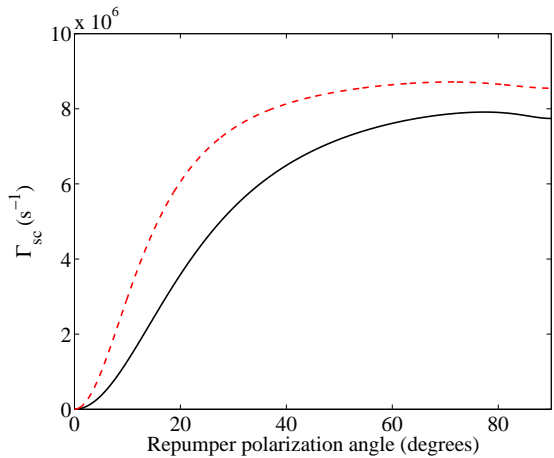}%
\caption{\label{fig:RepumpAngle}(Color online)  Scattering rate as a function of the angle between the magnetic field and the linear repumper polarization when dark states are destabilized using $B=50\;\mu$T [solid (black) curve] and $B=100\;\mu$T [dashed (red) curve]. $\Omega_\mathrm{c} = \Omega_\mathrm{r} = \Gamma$, $\delta_\mathrm{c} = -\Gamma/2$, $\delta_\mathrm{r} = \Gamma/2$, $\gamma_{g,m}=\Gamma/10$, and cooling laser $\sigma$ polarized.}
\end{figure}

The polarization angle of the repumper is crucial for the magnetic-field destabilization to work (see Fig.~\ref{fig:RepumpAngle}). It works only when the dark states are superpositions, i.e., when the repumper contains both $\sigma^+$ and $\sigma^-$. For all other polarizations ($\pi$, $\sigma^+$, $\sigma^-$, combination of $\pi$ and $\sigma^+$ \emph{or} $\sigma^-$, but not both) there is at least one dark eigenstate into which optical pumping occurs.

\section{Polarization modulation \label{sec:polarizationmodulation}}

Dark state destabilization using an external magnetic field has been discussed in Sec.~\ref{sec:dark_states}. However, there are applications where a strong magnetic field is not desired, in particular ion clocks, where typically fields of only a few microteslas are used.

A simple and commonly used way to modulate the repumper polarization is to pass the beam through an electro-optical modulator (EOM) at a 45$^\circ$ angle to the optical axis. The resulting polarization (with beam direction as QA) is
\begin{eqnarray}
\mathbf{u}_\text{EOM}(t) &=& \frac{1}{2}\left[(1+i e^{-i\varphi_\text{EOM}(t)}) \mathbf{u}_{+1} + (1-i e^{-i\varphi_\text{EOM}(t)}) \mathbf{u}_{-1} \right] \nonumber \\
&=& -\frac{i}{\sqrt{2}} (e^{-i\varphi_\text{EOM}(t)} \mathbf{u}_x + \mathbf{u}_y). \label{eq:EOM-pol}
\end{eqnarray}
where the phase retardation is \cite{Berkeland2002a}
\begin{equation} \label{eq:EOM-ret}
\varphi_\text{EOM}(t) = \frac{1}{2} \Phi_\text{EOM} ( 1-\cos{\omega_\text{EOM}t}).
\end{equation}
If the modulation frequency $\omega_\text{EOM}=0$ or the modulation amplitude $\Phi_\text{EOM}=0$, the resulting polarization is the stationary linear polarization incident on the EOM, $(\mathbf{u}_x + \mathbf{u}_y)/\sqrt{2}$.
In the following, we consider only the EOM modulation technique. Modulation techniques using acousto-optic modulators (AOMs) give similar results \cite{Berkeland2002a}.

\begin{figure}[h]
\includegraphics[width=.85\columnwidth]{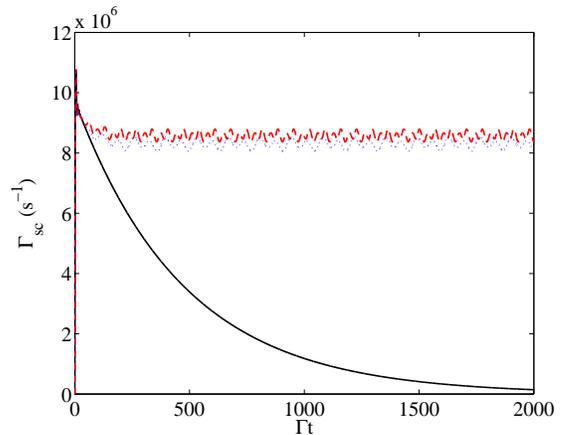}%
\caption{\label{fig:pol-mod}(Color online) Scattering rate for stationary $\sigma$ repumper [solid (black) curve] and polarization modulated according to Eqs.~(\ref{eq:EOM-pol}) and (\ref{eq:EOM-ret}) at $\omega_\text{EOM}=2\pi\times 2\;\text{MHz} \approx 0.09 \Gamma$ [dashed (red) curve]. Modulation at 1\;MHz plus a 50-$\mu$T field along $\mathbf{u}_x$ gives approximately the same result [dotted (blue) curve]. $\Phi_\text{EOM}=\pi$, $\Omega_\mathrm{c} = \Omega_\mathrm{r} = \Gamma$, $\delta_\mathrm{c} = -\Gamma/2$, $\delta_\mathrm{r} = \Gamma/2$, $\gamma_{g,m}=\Gamma/10$, and cooling laser $\mathbf{u}_x$ polarized.}
\end{figure}

Figure~\ref{fig:pol-mod} shows that the dark state destabilization due to polarization modulation is very similar to magnetic field destabilization when the dark state evolution rate is the same (cf.\ Fig.~\ref{fig:m-states}), except for some oscillations at the modulation frequency $\omega_\text{EOM}$. The two effects also sum up almost fully [see the dotted (blue) curve in Fig.~\ref{fig:pol-mod}].

\begin{figure}[h]
\includegraphics[width=1\columnwidth]{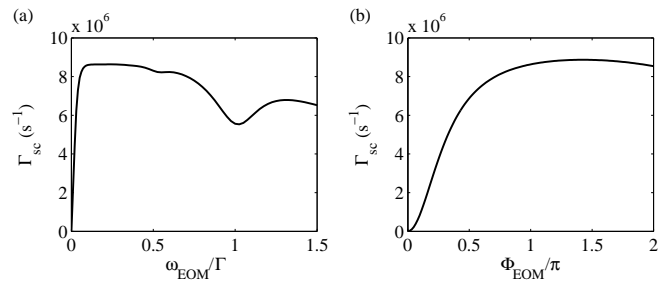}%
\caption{\label{fig:EOM} Quasi-steady-state scattering rate (a) as a function of modulation frequency $\omega_\text{EOM}$ for $\Phi_\text{EOM}=\pi$ and (b) as a function of modulation amplitude $\Phi_\text{EOM}$ for $\omega_\text{EOM}=\Gamma/4$. $\Omega_\mathrm{c} = \Omega_\mathrm{r} = \Gamma$, $\delta_\mathrm{c} = -\Gamma/2$, $\delta_\mathrm{r} = \Gamma/2$, $\gamma_{g,m}=\Gamma/10$, cooling laser $\mathbf{u}_x$ polarized, and repumper as in Eqs.~(\ref{eq:EOM-pol}) and (\ref{eq:EOM-ret}).}
\end{figure}

Figure~\ref{fig:EOM}(a) shows the quasi-steady-state scattering rate as a function of the modulation frequency $\omega_\text{EOM}$. As shown in Fig.~\ref{fig:pol-mod}, the dark-state destabilization is efficient already at $\omega_\text{EOM}\approx 0.1 \Gamma$ for the parameters used here. As expected, there are dips at $\Gamma$ and $\Gamma/2$, where the first- and second-order red sidebands of the repumper are at Raman resonance. Assuming we have a fixed EOM modulation frequency, this must be taken into account when choosing the repumper detuning. The dips become more pronounced if the Rabi frequencies are increased or if $\gamma_{g,m}$ is decreased. Figure~\ref{fig:EOM}(b) shows that for $\omega_\text{EOM}=\Gamma/4$, the scattering rate is maximized for the modulation amplitude $\Phi_\text{EOM} \approx 1.4\pi$. At lower amplitudes the polarization variation is not sufficient to fully destabilize the dark states, and at higher amplitudes an increasing part of the repumper intensity is transferred into sidebands further from resonance.

\section{Ion motion}

The ion undergoes both secular (thermal) motion at the trap frequencies and micromotion at the applied trap drive frequency. If the ion is not significantly perturbed from the RF voltage symmetry node of the trap by stray fields so that the micromotion amplitude is kept low, the Doppler cooling is not compromised and the ion experiences an approximately constant laser intensity near the beam focus. Under these circumstances, the motion can be accounted for by considering the rest frame of the ion and including the oscillations in the phases of the laser fields.

The equations of motion of a single ion in a quadrupole trap are given by the Mathieu equation and the first-order solution in the trap coordinate system $XYZ = R_XR_YR_Z$ is \cite{Berkeland1998a}
\begin{equation}
R_i(t) \approx \left[ R_{0i} + R_{\text{sec},i} \cos{(\omega_{\text{sec},i} t + \theta_{\text{sec},i})} \right] \left( 1+ \frac{q_i}{2} \cos{\Omega t} \right).
\end{equation}
Here $R_{\text{sec},i}$ is the secular motion amplitude, which can be obtained from the temperature of the ion, $E_{\mathrm{K}i} = k_\mathrm{B} T_i \approx \frac{1}{2} m R_{\text{sec},i}^2 \omega_{\text{sec},i}^2$ \cite{Berkeland1998a},
$\omega_{\text{sec},i}$ is the trap frequency, and $\theta_{\text{sec},i}$ is the phase, determined by the initial position and momentum of the ion. The trap drive frequency is $\Omega$ and $q_i$ is a parameter that depends on the trap geometry, drive voltage, and ion mass. $R_{0i}$ is the displacement of the average position of the ion that is caused by static electric fields and that results in excess micromotion.

We consider a geometry, where the endcap trap $Z$ axis is at an angle $\alpha = 35.3^\circ$ to the horizontal plane. The possible beam directions are vertical ($z$) and horizontal at $\pm 45^\circ$ to the horizontal projection of the trap axis ($x$ and $y$). In the beam coordinate system, the ion motion is then given by
\begin{subequations}
\begin{eqnarray}
r_{x}(t) &=& \frac{1}{\sqrt{2}} \left[ R_{X}(t) -\sin{\alpha} \,R_{Y}(t) -\cos{\alpha} \,R_{Z}(t) \right], \\
r_{y}(t) &=& \frac{1}{\sqrt{2}} \left[ R_{X}(t) +\sin{\alpha} \,R_{Y}(t) +\cos{\alpha} \,R_{Z}(t) \right], \\
r_{z}(t) &=& -\cos{\alpha} \,R_{Y}(t) +\sin{\alpha} \,R_{Z}(t).
\end{eqnarray}
\end{subequations}
This geometry is chosen to make the projection of each beam direction on the trap $Z$ axis equal in magnitude ($\cos{\alpha}/\sqrt{2} = \sin{\alpha}$). In the rest frame of the ion, the phase of a laser beam traveling along the $\mathbf{u}_l$ axis is then given by $\exp{\{-i[\omega_j t - k_jr_l(t) - \phi_{jl}]\}}$ ($j=\mathrm{c},\mathrm{r}$). The phase $\phi_{jl}$ is not relevant.

\begin{figure}[h]
\includegraphics[width=.85\columnwidth]{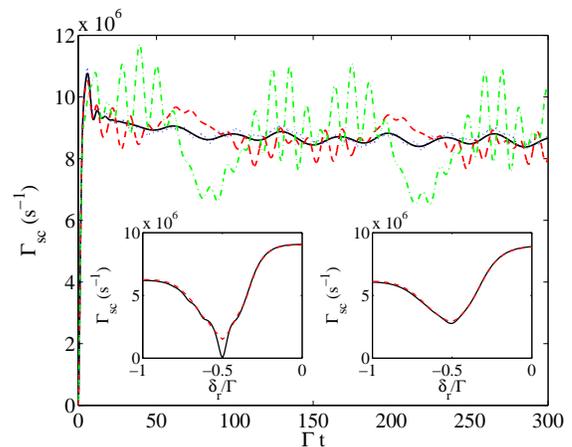}%
\caption{\label{fig:ionmotion}(Color online) Scattering rate for a stationary ion [$T=0$; solid (black) curve] and at temperatures $0.5$\;mK [dashed (red) curve] and $5$\;mK [dashed-dotted (green) curve]. The average over 250 curves with $T=5$\;mK and random secular motion phases $\theta_i$ approaches the stationary case [dotted (blue) curve]. Left inset: CPT resonances for a stationary ion [solid (black) curve] and at $T=5$\;mK [dashed (red) curve] for $\gamma_{g,m}=0$. Right inset: Corresponding CPT resonances for $\gamma_{g,m}=0.04\Gamma$.
Parameters not mentioned are the same as for the dashed (red) curve in Fig.~\ref{fig:pol-mod}.}
\end{figure}

Figure~\ref{fig:ionmotion} illustrates the effect of ion motion using numerical parameters typical for an endcap trap: $q_Z=-2q_{X,Y}=0.4$, $\omega_{\text{sec},Z}=2\omega_{\text{sec},X,Y}=2\pi\times 2$\;MHz, and $\Omega=2\pi\times 12$\;MHz. The solid (black) curve is the scattering rate for a stationary ion, $T=0$, equal to the dashed (red) curve in Fig.~\ref{fig:pol-mod} (note the different horizontal axis), and shows oscillations at the frequency $\omega_\mathrm{EOM}$ only. At finite temperatures, there are additional oscillations at the trap frequencies $\omega_{\text{sec},X,Y}$ and $\omega_{\text{sec},Z}$ and at the drive frequency $\Omega$, and the time dependence of the scattering rate is rather complex. The oscillation amplitudes increase with increasing temperature; see the dashed (red) and dash-dotted (green) curves in Fig.~\ref{fig:ionmotion}, which correspond to temperatures of $0.5$\;mK (Doppler limit for $^{88}\mathrm{Sr}^+$) and $5$\;mK, respectively. However, as the density matrix formalism describes an ensemble average over the internal degrees of freedom, it is not physically meaningful to select fixed values for the external degrees of freedom, i.e., the secular motion phases $\theta_i$ that depend on the initial position and velocity of the ion. Instead, we must average over a large number of solutions with random phases $\theta_i$. The dotted (blue) curve in Fig.~\ref{fig:ionmotion} shows that an average over 250 solutions with $T=5$\;mK is very close to the curve for a stationary ion.

Intuitively, one expects the ion motion to have the most significant effect in the vicinity of a narrow CPT resonance. The left-hand inset in Fig.~\ref{fig:ionmotion} shows CPT resonances for $\gamma_{g,m}=0$. For a stationary ion (solid black curve), the scattering rate vanishes at CPT resonance, whereas the resonance is significantly ``rounded off'' in the $T=5$\;mK case (dashed red curve). However, the right-hand inset in Fig.~\ref{fig:ionmotion} shows the corresponding curves for $\gamma_{g,m}=0.04\Gamma$ (the value that is used in the comparison with experiments in Sec.~\ref{sec:exp}) and now the difference between the curves is minute. For the inset curves, the ensemble average was evaluated as a time average once quasi-steady-state was reached, which was particularly convenient as the involved frequencies were chosen to be multiples of $\omega_{\text{sec},X,Y}$.

Thus, unless one is studying CPT resonances with a very low coherence dephasing rate $\gamma_{g,m}$, one can closely approximate the scattering rate without accounting for ion motion in cases where the ion motion is of a low amplitude and micromotion effects do not dominate.

\section{Comparison with experiments \label{sec:exp}}

\subsection{Endcap trap setup}

We now compare the theoretical results to experimental spectra measured using the $^{88}\mathrm{Sr}^+$ endcap trap at NRC \cite{Dube2010a}. Briefly, the experiment employs a single ion of $^{88}\mathrm{Sr}^+$ held in an endcap trap \cite{Schrama1993a,Sinclair2001a}. The trap characteristic dimensions have endcap electrodes of radius $r_0 = 0.25$\;mm with separation $2z_0 = 0.541\pm 0.015$\;mm made of molybdenum. The shield electrodes were kept at ground potential, had a diameter of $2r_2=2$\;mm, and were made from tantalum tubing. The entire trap structure was composed of nonmagnetic materials and housed in an ultra-high-vacuum environment, where the background pressure was 10\;nPa, composed primarily of H$_2$. At these pressures, the mean collision times are of the order of $10^3$\;s \cite{Madej1990a} and do not contribute to the dephasing rates of the observed line shapes. The ion trap is operated at a trap frequency of $\Omega = 2\pi \times 14.4$\;MHz with a voltage amplitude of $V_0 = 200$\;V. The trap secular frequencies were $\omega_{\text{sec},Z}/2\pi = 2.3$\;MHz and $\omega_{\text{sec},X,Y}/2\pi = 1.2$\;MHz in the radial direction, with the difference between the two radial frequencies being about 20\;kHz. The ion is loaded into the trap using a weak effusive source of Sr atoms and photoionized using 461- and 404-nm laser beams which are turned off following the loading process \cite{Brownnutt2007a}. In this way, micromotion due to patch potentials of deposited Sr is reduced to a minimum and only small adjustments of the trap compensation electrodes are necessary between loadings.  The trap is enclosed in a double-layer magnetic shield and the residual field at the trap center is $\mathbf{B} = (-1.22 \mathbf{u}_x -0.70 \mathbf{u}_y + 1.47 \mathbf{u}_z) \;\mu\mathrm{T}$ ($|\mathbf{B}|=2.03 \;\mu\mathrm{T}$).

In the experiment, both beams propagate in the $\mathbf{u}_y$ direction. The cooling laser is linearly polarized in the $\mathbf{u}_z$ direction, whereas the repumper polarization is modulated like $\mathbf{u}_\mathrm{r} = 2^{-1/2}(\mathbf{u}_x + e^{-i\varphi_\text{EOM}(t)} \mathbf{u}_z)$. The EOM frequency is 12\;MHz ($\omega_\text{EOM} = 0.556 \Gamma$) and the amplitude is $\Phi_\text{EOM} = (1.1\pm0.2) \pi$. Typical beam parameters at trap center are $w_\mathrm{c} = 16\pm 1\;\mu$m and $w_\mathrm{r} = 50.5 \pm 2 \;\mu$m. The linewidth of the repumper (a diode pumped fiber laser) is specified to below 10\;kHz and can be neglected. The repump laser is stabilized in its drift by stabilization to a transfer optical cavity whose length is controlled using a 633-nm polarization-stabilized HeNe laser system which maintains the repumper laser within 2\;MHz over extended periods of time. The cooling laser system has been described in a previous publication \cite{Shiner2007a}. The laser is a 422-nm diode laser system which is stabilized in its short-term frequency fluctuations to the side of a transmission fringe of an optical cavity and then controlled in its long-term drift by locking to the saturated absorption of the $5s \,^2S_{1/2} - 6p \,^2P_{1/2}$ line in $^{85}\mathrm{Rb}$.
The linewidth of the cooling laser has been determined by two different independent techniques to be $2.4\pm 0.8$\;MHz \cite{Shiner2007a} and $0.6\pm 0.3$\;MHz. A ground-metastable coherence decay rate $\gamma_{g,m} = 0.04\Gamma$, corresponding to a linewidth of 1.7\;MHz, is used in the current comparison, as it gives the best agreement between experiments and theory. The cooling laser is frequency referenced to the $^{85}\mathrm{Rb}$ $5s \,^2S_{1/2}\, (F=2) - 6p \,^2P_{1/2}\, (F=3)$ transition  \cite{Madej1998a, Shiner2007a}, which provides the absolute frequency scale for the cooling laser line shapes. In the repumper line shapes, the absolute frequency scale is determined from the position of the CPT resonances.

The laser Rabi frequencies are determined by measuring the beam powers and waists, giving an uncertainty of approximately 20--25\%. However, the slightest misalignment of the beams affects the Rabi frequencies experienced by the ion, so the actual uncertainty is larger. The ion trap fluorescence is optimized at low laser intensities with neutral density filters in the beam paths. The filters are then removed for the measurements at high Rabi frequencies and it is possible that this alters the beam alignment so that the true Rabi frequencies are lower than estimated in these cases.

Ion fluorescence is detected by an $f = 27$\;mm ($f$-number $f/\#=0.9$) aspheric collection lens located in the vacuum chamber which focuses the light onto a pinhole in front of a photon counting photomultiplier system. Detected count rates at optimum fluorescence are typically at $10\,000$ cps. The detected fluorescence rate is sensitive to scattered cooling laser light, mainly from the trap electrodes. This background has not been removed from the experimental data. However, the level of parasitic light in the current experiment is typically less than $1/50$ of the observed single ion fluorescence under optimum conditions. The relation between the detected fluorescence rate and the total scattering rate is not known and the experimental data are hence given in arbitrary units. A second viewport opposite the photon counting photomultiplier has a commercial camera lens system ($f/\#= 4$) which images the trap central region on a photon counting CCD camera system. This allows the initial optimization of beam alignment and ion fluorescence to be performed. In addition, it is used as a first stage of minimizing the micromotion using external compensation electrodes in the radial plane and axial dc bias of the endcap electrodes in the axial direction. Final optimization of the ion micromotion is achieved by the measurement and optimization of the observed sideband features in the spectra of the ion probed on the reference ion clock transition at 445\;THz. In this way the total Stark and time dilation perturbations of the single ion are maintained below the $2\times 10^{-17}$ fractional shift for the reference transition at 445\;THz and micromotion effects are deemed negligible for the current line-shape scans. In addition, scans of the ion secular motion sidebands have placed an operating temperature of $2\pm 1$\;mK for all trap canonical directions under optimum laser cooling, thus well satisfying the condition for Doppler free spectra of the ion line shapes. Departures from the ideal low ion kinetic temperatures are anticipated when the laser powers cause low fluorescence scattering and detunings are near the ion line center.

\subsection{Results}

\begin{figure}[h]
\includegraphics[width=.85\columnwidth]{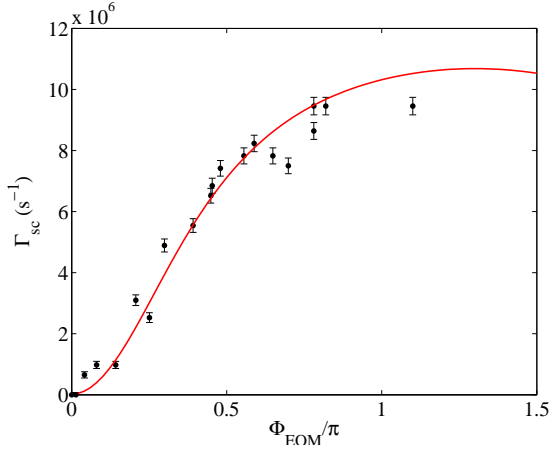}%
\caption{\label{fig:ModAmp}(Color online) Measured [filled (black) circles] and calculated [solid (red) curve] scattering rate as a function of EOM modulation amplitude. $\Omega_\mathrm{c}/\Gamma = 1.7$, $\Omega_\mathrm{r}/\Gamma = 0.77$, $\delta_\mathrm{c}/\Gamma = -0.74$, $\delta_\mathrm{r}/\Gamma = 0.7$.}
\end{figure}

Figure \ref{fig:ModAmp} shows the scattering rate as a function of the EOM modulation amplitude as in Fig.~\ref{fig:EOM}(b). The only adjusted parameter in the comparison was the ratio of the observed detected count rate to the calculated absolute scattering rate, which was set at 8150. Other parameters were based on the measured experimental conditions. As shown, the observed dependence of fluorescence matches the calculated relation very well confirming optimum fluorescence at $\Phi_\text{EOM}/\pi > 1$.

\begin{figure}[h]
\includegraphics[width=.9\columnwidth]{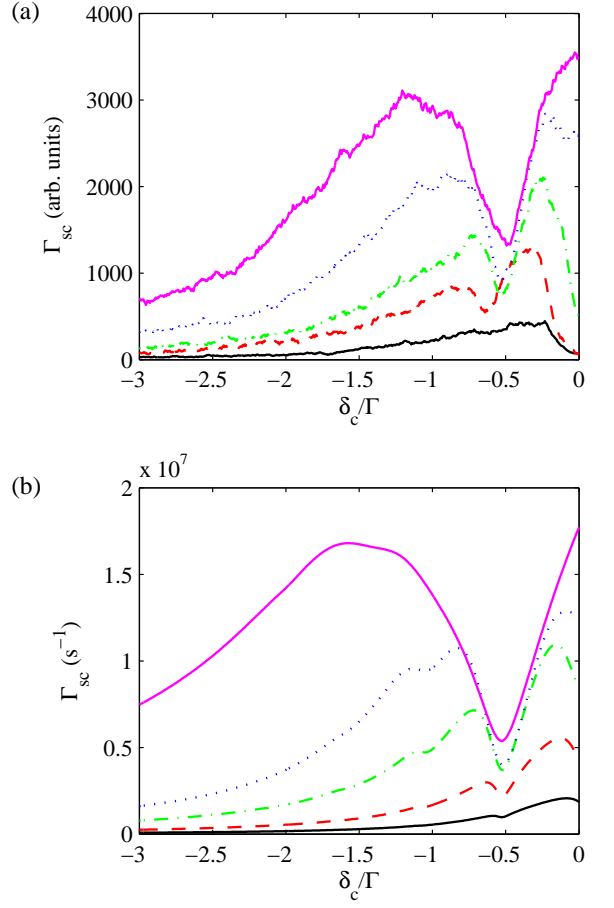}%
\caption{\label{fig:SPscans}(Color online) (a) Measured and (b) calculated scattering rate as a function of cooling laser detuning for different Rabi frequency pairs $[\Omega_\mathrm{c}/\Gamma, \,\Omega_\mathrm{r}/\Gamma]$. From bottom to top: $[0.36,\,0.39]$ [solid (black) curve], $[0.64,\,0.70]$ [dashed (red) curve], $[1.13,\,1.25]$ [dash-dotted (green) curve], $[1.60,\,2.22]$ [dotted (blue) curve], and $[3.58,\,3.22]$ [solid (magenta) curve]. $\delta_\mathrm{r}/\Gamma = -0.53$.}
\end{figure}

\begin{figure}[h]
\includegraphics[width=1\columnwidth]{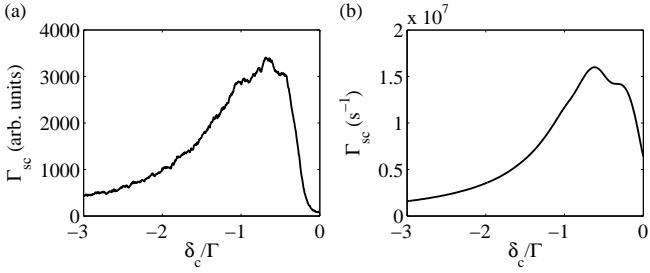}%
\caption{\label{fig:SPnormal} (a) Measured and (b) calculated scattering rate as a function of cooling laser detuning for $\delta_\mathrm{r}/\Gamma = 0.09$. $\Omega_\mathrm{c}/\Gamma = 1.60$, $\Omega_\mathrm{r}/\Gamma = 2.22$.}
\end{figure}

Figure \ref{fig:SPscans} shows line shapes obtained by scanning the cooling laser detuning for increasing  Rabi frequencies. The repumper was red-detuned in order to make the CPT resonances visible. Only the red half of the spectrum can be measured and in some of the experimental curves one can see how the scattering rate drops close to $\delta_\mathrm{c} = 0$ as the laser cooling is compromised. The agreement between the experimental and the calculated line shapes is very good, except for the highest Rabi frequency line shape. In this case, both the optical line and the CPT resonance are considerably more power broadened in the calculated curve, indicating that the experimental Rabi frequencies are lower than estimated. This is probably due to beam misalignment as mentioned above. In the second lowest [dashed (red)] curve, the CPT resonance is offset, as the repumper detuning has drifted slightly. Figure~\ref{fig:SPnormal} shows the corresponding line shapes for more normal operating conditions with a repumper detuning of $0.09\Gamma$. Again, lower fluorescence levels are observed near line center, presumably due to ion heating. If a narrower linewidth is desired, even lower Rabi frequencies have to be used at the sacrifice of the ion scattering rate.

\begin{figure}[h]
\includegraphics[width=.9\columnwidth]{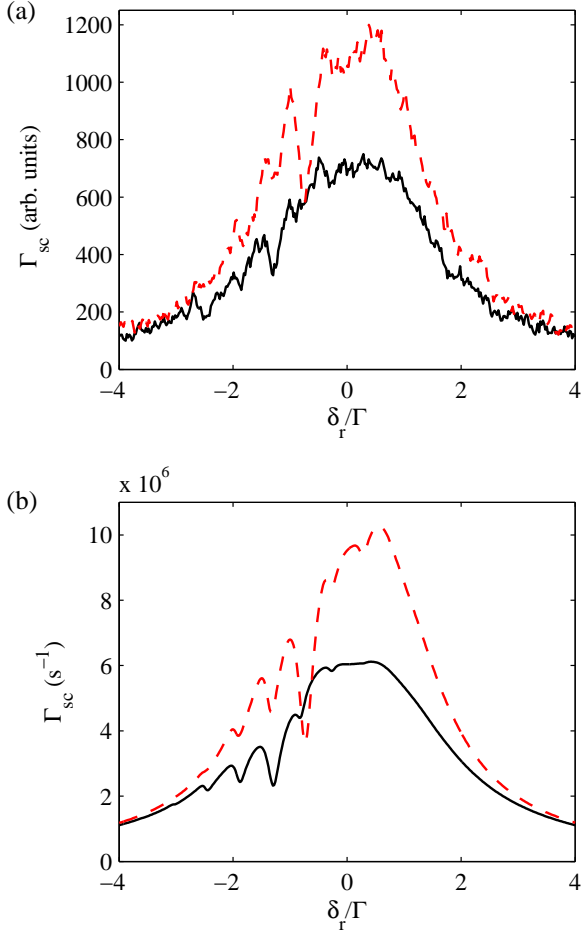}%
\caption{\label{fig:PDdc}(Color online) (a) Measured and (b) calculated scattering rate as a function of repumper detuning for two cooling detunings: $\delta_\mathrm{c}/\Gamma = -1.30$ [solid (black) curve]  and $\delta_\mathrm{c}/\Gamma = -0.74$ [dashed (red) curve]. $\Omega_\mathrm{c}/\Gamma = 1.69$, $\Omega_\mathrm{r}/\Gamma = 0.72$.}
\end{figure}

Figure \ref{fig:PDdc} shows line shapes obtained by scanning the repumper frequency for two cooling laser detunings. Again, the agreement is very good considering the uncertainties of the experimental parameters. The multiple coherence dips in the spectra arise from the different orders of the EOM frequency falling in Raman (CPT) resonance. Again it should be noted that a significant change in the observed line shape is seen for small changes in cooling detuning which is well matched by the calculations.

\begin{figure}[h]
\includegraphics[width=.9\columnwidth]{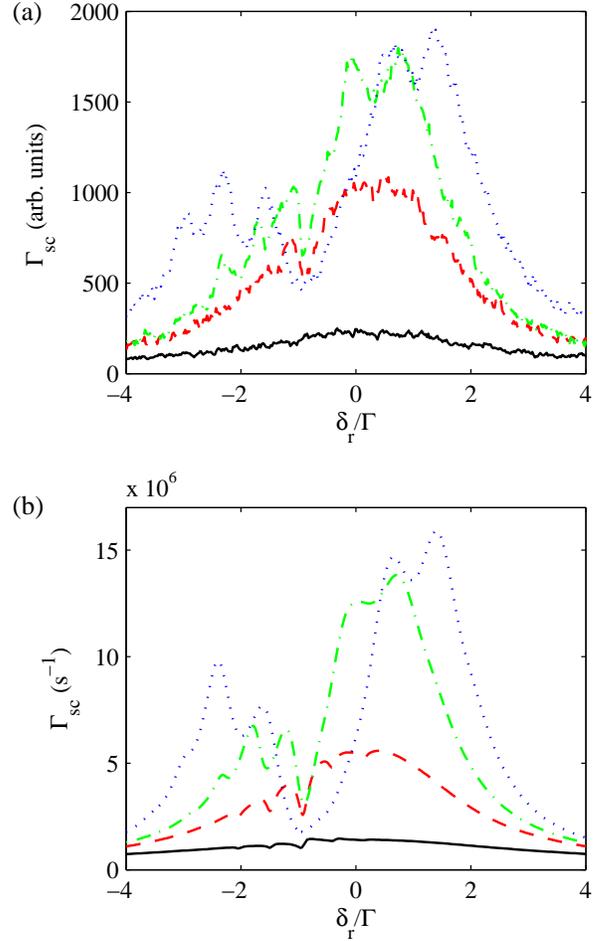}%
\caption{\label{fig:PDRc}(Color online) (a) Measured and (b) calculated scattering rate as a function of repumper detuning for increasing cooling Rabi frequencies $\Omega_\mathrm{c}/\Gamma$: $0.53$ [solid (black) curve], $1.19$ [dashed (red) curve], $3.00$ [dash-dotted (green) curve], and $6.71$ [dotted (blue) curve]. $\delta_\mathrm{c}/\Gamma = -0.92$, $\Omega_\mathrm{r}/\Gamma = 0.72$.}
\end{figure}

\begin{figure}[h]
\includegraphics[width=.75\columnwidth]{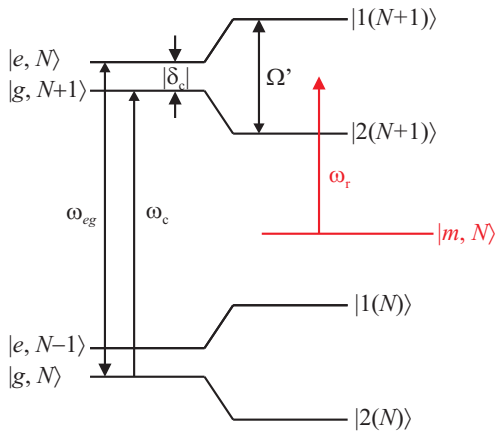}%
\caption{\label{fig:DressedStates}(Color online) Dressed state model: uncoupled levels at the left and dressed states at the right.}
\end{figure}

Figure \ref{fig:PDRc} shows repumper line shapes for increasing cooling laser Rabi frequencies.
The two-peak spectrum that occurs for the highest $\Omega_\mathrm{c}$ values can be explained using dressed states \cite{Cohen-Tannoudji:API}. The dressed states are the eigenstates of the interacting ion ($|g\rangle$ and $|e\rangle$ levels) plus cooling laser system when the laser is strong ($C_{82} \Omega_\mathrm{c} \gg \Gamma$). The left-hand side of Fig.~\ref{fig:DressedStates} shows the uncoupled states of the ion-photon system. The energy level scheme consists of a ladder of manifolds, separated by the cooling laser frequency $\omega_\mathrm{c}$, two of which are shown in the figure. Each manifold consists of two closely spaced ($|\delta_\mathrm{c}|$) states $|g,N\rangle$ and $|e,N-1\rangle$, where the first letter refers to the state of the ion and the second to the number of cooling laser photons. When the ion-laser interaction is taken into account, the strong cooling laser couples the ground and excited states and the new eigenstates are the two dressed states $|1(N)\rangle$ and $|2(N)\rangle$ shown at the right in Fig.~\ref{fig:DressedStates}. These are both superpositions of $|g\rangle$ and $|e\rangle$ and are separated by the generalized Rabi frequency $\Omega' = \sqrt{C_{82}^2 \Omega_\mathrm{c}^2 + \delta_\mathrm{c}^2}$. The frequencies of the two dressed states in relation to the unperturbed excited state are $\Delta\omega_{1,2} = (\delta_\mathrm{c} \pm \Omega')/2$. Since the $|g\rangle \rightarrow |e\rangle$ system is not closed, the population will be optically pumped into $|m\rangle$ except when the repumper is resonant with one of the dressed states. For the highest $\Omega_\mathrm{c}$ line shape in Fig.~\ref{fig:PDRc}, the dressed state frequencies are $\Delta\omega_1 = 1.0\Gamma$ and $\Delta\omega_2 = -1.9\Gamma$.
Thus the simple dressed state model predicts the peak positions very accurately. In addition, each dressed state peak has a CPT resonance at the very center. In practice, such high applied cooling intensities should not be used if low ion kinetic temperatures are desired, as the high intensities induce power broadening and a reduction in the obtainable minimum laser cooling temperatures \cite{Javanainen1980a}.

\begin{figure}[h]
\includegraphics[width=1\columnwidth]{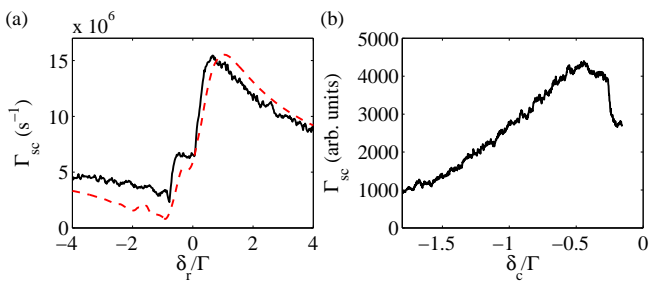}%
\caption{\label{fig:BrightNormal}(Color online) (a) Measured [solid (black) curve] and calculated [dashed (red) curve] scattering rate as a function of repumper detuning for a high repumper Rabi frequency $\Omega_\mathrm{r}/\Gamma = 5.90$. $\delta_\mathrm{c}/\Gamma = -0.92$, $\Omega_\mathrm{c}/\Gamma = 1.19$. The measured curve was multiplied by a factor of 5150 to make the amplitudes equal. (b) Scattering rate as a function of cooling laser detuning for typical experimental conditions: $\Omega_\mathrm{c}/\Gamma = 1.19$, $\Omega_\mathrm{r}/\Gamma = 1.8$, and repumper blue-detuned.}
\end{figure}

If instead the repumper has a relatively high Rabi frequency [Fig.~\ref{fig:BrightNormal}(a)], a dispersive line shape is obtained, as confirmed in the experimental spectrum. The ``bright lateral part'' next to the dark resonance was first observed by Alzetta \emph{et al}.\ \cite{Alzetta1979a} and occurs when the two Rabi frequencies are considerably different and the detunings are non-zero.  It is caused by stimulated Raman scattering between the ground and the metastable states \cite{Lounis1992a} and has been studied also in single trapped ions \cite{Janik1985a,Siemers1992a}.

Figure~\ref{fig:BrightNormal}(b) shows a cooling laser line shape for typical trap operation parameters. The repumper is blue-detuned, but the exact detuning is not known. Strong ion fluorescence is obtained with the linewidth approaching the natural width value. The FWHM of the line is difficult to determine exactly from such a half spectrum with possible changes in ion fluorescence near line center, but can be seen to be $\lesssim 2\Gamma$.


\section{Conclusions}

In this paper, we have studied the different types of dark states that can occur in a single trapped $^{88}\mathrm{Sr}^+$ ion due to the cooling and repumping laser fields. Efficient Doppler cooling and fluorescence detection of the ion require a high scattering rate, which is only achieved if dark states are prevented from forming. Dark states within the $5s\,^2S_{1/2}$ ground state can be prevented simply by using a linearly polarized cooling beam. Dark states between $5s\,^2S_{1/2}$ ground state and $4d\,^2D_{3/2}$ metastable state sublevels can be tuned away by using a positive repumper detuning (assuming that the cooling laser detuning is negative for Doppler cooling) when the lasers have narrow lines and modest intensities so that the CPT resonances are not heavily broadened. On the other hand, the cross-correlation between the two lasers and their linewidths affect the dephasing rate of the ground-metastable coherences. This means that if the two lasers are uncorrelated and the sum of their linewidths is of the order of the natural linewidth of the ion, the contrast of the ground-metastable dark states will vanish. Dark states within the metastable state are thus the main problem. These can be destabilized using a magnetic field or, in applications where strong fields cannot be tolerated such as ion clocks, by modulating the repumper polarization.

Our results regarding the $SP$ linewidth and optimum parameters essentially agree with those of Berkeland and Boshier \cite{Berkeland2002a}. Power broadening by the cooling laser becomes significant for $\Omega_\mathrm{c}/\Gamma > 1/\sqrt{3} \approx 0.6$. However, for many practical applications it might be desirable to increase the Rabi frequency to $\Omega_\mathrm{c}/\Gamma \approx 1$, which increases the scattering rate by more than a factor of 2, while the linewidth increases only by approximately 15\%.
If the repumper Rabi frequency is lower than the cooling laser Rabi frequency, there is significant broadening due to optical pumping into the metastable state. On the other hand, both lasers cause power broadening of the CPT resonances. A too high $\Omega_\mathrm{r}$ value causes the CPT resonance to extend to the red side of the line. Depending on the exact values of $\Omega_\mathrm{c}$ and $\delta_\mathrm{r}$, a repumper Rabi frequency of $\Omega_\mathrm{r} \approx (1\ldots1.5)\Omega_\mathrm{c}$ gives the maximum scattering rate, while the CPT resonances remain sufficiently narrow that they can be tuned away by a positive repumper detuning $\delta_\mathrm{r} \approx \Gamma/2$.

We have also analyzed the effect of ion motion for cases where its amplitude is low, i.e., the ion is sufficiently cooled and does not undergo excess micromotion. We found that at a $5$\;mK temperature, high-contrast CPT resonances, associated with a low ground-metastable coherence dephasing rate $\gamma_{g,m}$, i.e., very narrow linewidth or cross-correlated lasers, are rounded off by the ion motion. For higher dephasing rates $\gamma_{g,m} \gtrsim 0.04$, at lower temperatures, or further from CPT resonance, the effect of ion motion can be neglected.

The theoretical results have been compared to experimental data from the $^{88}\mathrm{Sr}^+$ endcap trap at the National Research Council of Canada and the agreement was found to be very good considering the uncertainties of the experimental parameters. The results confirm that the system can be well described by the current theoretical framework and can shed light on the full optimization of single ion fluorescence and laser cooling for single trapped ions applied to atomic frequency standards or test systems for quantum manipulation studies. The rich variety of observed behaviors can now be understood and optimal parameters chosen depending on the desired operating conditions needed. It is anticipated that these results will certainly aid in providing the optimal high fluorescence rates needed for efficient detection of single ions and the low kinetic temperatures from laser cooling needed for a perturbation free nearly isolated quantum system.


\begin{acknowledgments}
The authors would like to thank P.~Dub\'e for essential aid in the construction and operation of the NRC single-ion trap system and Z.~Zhou for assistance in portions of the data acquisition. This work received partial support from the Natural Sciences and Engineering Research Council (NSERC).
The work at MIKES was funded by the Academy of Finland (Project No.\ 138894).
\end{acknowledgments}


%

\end{document}